\documentclass[preprint2]{aastex}

\newcommand{\vsini} {$v$\,sin\,$i$}
\newcommand{\kms} {km\,s$^{-1}$}

\shorttitle{$Macroturbulent$ broadening and LPVs in B Supergiants}
\shortauthors{Sim\'on-D\'iaz et al.}

\begin{document}

\title{Observational evidence for a correlation between $macroturbulent$ broadening 
and line-profile variations in OB Supergiants\thanks{Based on observations made with the Nordic Optical Telescope, operated
on the island of La Palma jointly by Denmark, Finland, Iceland,
Norway, and Sweden, in the Spanish Observatorio del Roque de los
Muchachos of the Instituto de Astrofisica de Canarias.}}

\author{S. Sim\'on-D\'iaz}
\affil{Instituto de Astrof\'isica de Canarias, E-38200 La Laguna, Tenerife, Spain.}
\affil{Departamento de Astrof\'isica, Universidad de La Laguna, E-38205 La Laguna, Tenerife, Spain.}
\email{ssimon@iac.es}

\author{A. Herrero}
\affil{Instituto de Astrof\'isica de Canarias, E-38200 La Laguna, Tenerife, Spain.}
\affil{Departamento de Astrof\'isica, Universidad de La Laguna, E-38205 La Laguna, Tenerife, Spain.}

\author{K. Uytterhoeven}
\affil{Laboratoire AIM, CEA/DSM-CNRS-Universit\'e Paris Diderot; 
CEA, IRFU, SAp, centre de Saclay, F-91191, Gif-sur-Yvette, France.}

\author{N. Castro}
\affil{Instituto de Astrof\'isica de Canarias, E-38200 La Laguna, Tenerife, Spain.}
\affil{Departamento de Astrof\'isica, Universidad de La Laguna, E-38205 La Laguna, Tenerife, Spain.}

\author{C. Aerts}
\affil{Instituut voor Sterrenkunde, Katholieke Universiteit Leuven, Celestijnenlaan 200D, B-3001 Leuven, Belgium}
\affil{IMAPP, Department of Astrophysics, Radboud University Nijmegen, PO Box 9010, NL-6500 GL Nijmegen, the Netherlands}

\and

\author{J. Puls}
\affil{Universit\"atssternwarte M\"unchen, Scheinerstr. 1, D-81679 M\"unchen, Germany}

\begin{abstract}
The spectra of O and B supergiants are known to be affected by a
significant form of extra line broadening (usually referred to as {\it macroturbulence})
in addition to that produced by stellar rotation. Recent analyses of high resolution
spectra have shown that the interpretation of this line broadening as
a consequence of large scale turbulent motions would imply highly
supersonic velocity fields in photospheric regions, making this
scenario quite improbable. Stellar oscillations have been proposed as
a likely alternative explanation. As part of a long term observational
project, we are investigating the {macroturbulent}
broadening in O and B supergiants and its possible connection with
spectroscopic variability phenomena and stellar oscillations. In this
letter, we present the first encouraging results of our project, namely
firm observational evidence for a strong correlation between the
extra broadening and photospheric line-profile variations in a
sample of 13 supergiants with spectral types ranging from O9.5 to B8.
\end{abstract}

\keywords{stars: early-type --- stars: atmospheres --- stars: oscillations ---
          stars: rotation --- supergiants}

\section{Introduction}\label{intro}

The presence of an important extra line broadening mechanism (in addition to the rotational 
broadening and usually called $macroturbulence$) affecting the spectra of O and B supergiants 
(Sgs) was initially suggested by the deficit of narrow lined objects among these types 
of stars (e.g. \citealt{Sle56}; \citealt{Con77}; \citealt{How97}). The advent of 
high-quality spectra permitted confirmation 
 that  rotational broadening alone is insufficient to fit the line profiles 
in many objects, and to investigate the possible disentangling of both broadening 
contributions \citep[see e.g.][]{Rya02, Sim07}. 
These studies definitely showed that the effect of macroturbulence is dominant in 
the profiles of metal lines in early B\,Sgs.

Although it was named macroturbulence at some point, the interpretation of this extra broadening 
as the effect of turbulent motion is quite improbable. The effect is present in 
photospheric lines and affects the whole profile, even wavelengths close to the 
continuum. Therefore, whatever is producing the extra broadening has to be deeply 
rooted in the stellar photosphere (and possibly deeper), in layers where we do not 
expect any significant velocity field in these stars. 

\begin{figure}[t!]
   \centering
   \includegraphics[width=5.5cm, angle=90]{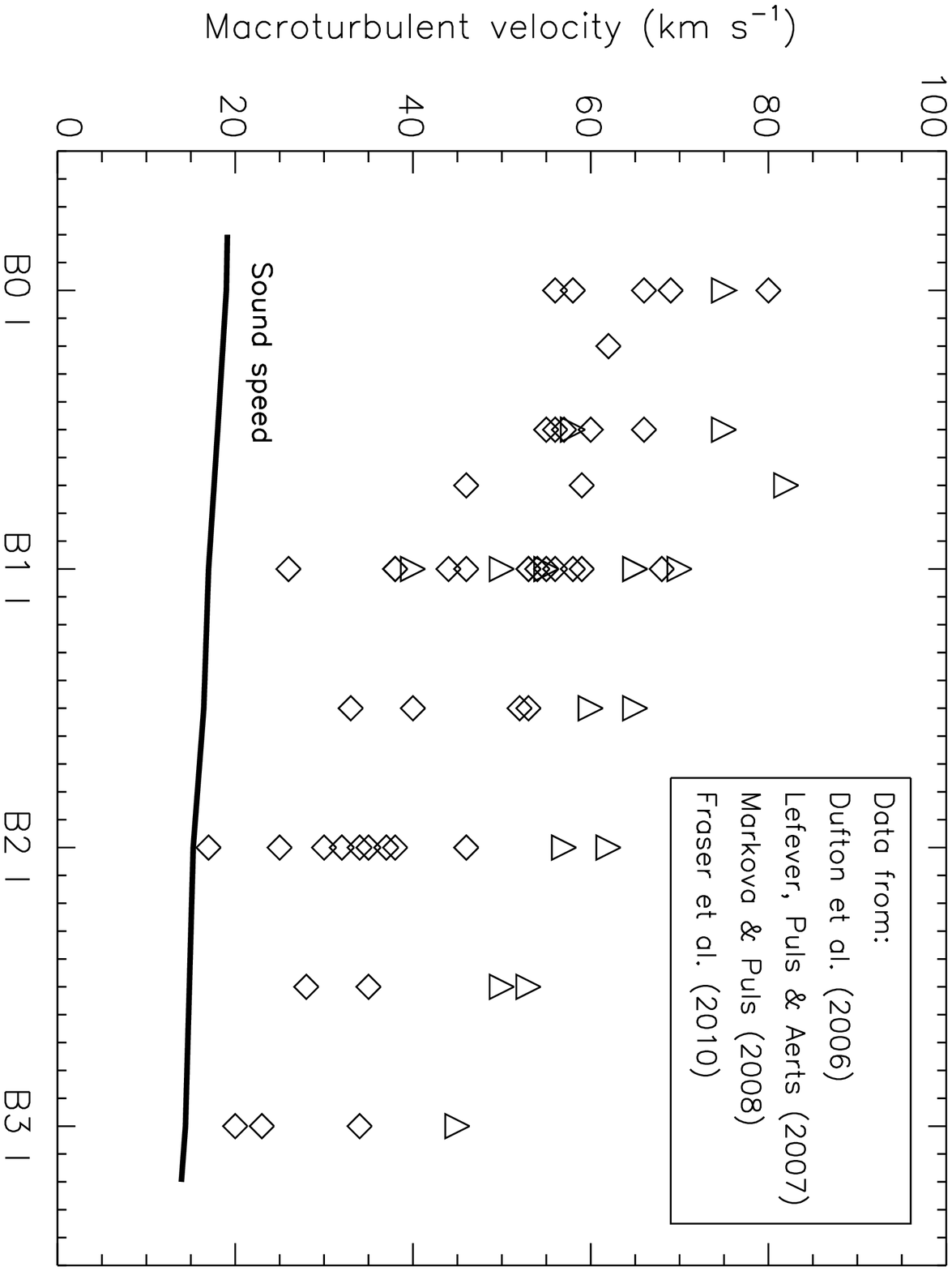}
      \caption{Macroturbulent velocities for B\,Sgs with spectral types B0-B3, 
      measured in recent studies. Two different symbols are used
      depending on the type of definition assumed for the macroturbulent profile. Diamonds: isotropic 
      Gaussian \citep{Duf06, Fra10}; triangles: radial-tangential definition \citep{Lef07, Mar08}.
      Characteristic values for the sound speed in the photospheres of these types of stars are also
      indicated as a solid line.}
         \label{f0}
   \end{figure}

If interpreted as turbulent motion, macroturbulence would 
represent {\em highly supersonic} velocities in many cases (see Figure \ref{f0}). This interpretation is 
incompatible with the previous statement.

   One physical mechanism suggested as the origin of this extra broadening
   relates to oscillations. Many OB\,Sgs are known to show photometric and
   spectroscopic variability.  Based on this, \citet{Luc76} postulated that this
   variability might be a pulsation phe\-no\-me\-non, and macroturbulence may be
   identified with the surface motions generated by the superposition of
   numerous non-radial oscillations.  More recently, \citet{Aer09} computed time
   series of line profiles for evolved massive stars broadened by rotation and
   thousands of low amplitude non-radial gravity mode oscillations and showed
   that the resulting profiles could mimic the observed ones.  In their
     paper, Aerts et al.\ (2009) considered the amplitudes of the pulsation
     modes to be sufficiently low to allow a linear superposition of mode
     velocities to derive the overall pulsational velocity eigenvector, i.e.,
     nonlinear mode coupling or other nonlinear effects were ignored. We refer
     the reader to Sect. 2.1 in \cite{Aer09} for a detailed description of the
     simulations, and a discussion about how the collective effect of low
     amplitude non-radial gravity modes can produce the inferred supersonic
     level of macroturbulent broadening when the oscillations are ignored in the
     interpretation of spectral lines.  Stellar oscillations are a plausible
   explanation for the extra broadening in O and B Sgs, but so far there is no
   direct observational evidence confirming this hypothesis.

   Two years ago we began an observational project aimed at investigating the
   macroturbulent broadening in O and B Sgs, and its possible connection with
   spectroscopic variability phenomena and stellar oscillations. In this letter,
   we present the first encouraging results, showing observational evidence for
   a strong correlation between this extra broadening and photospheric line
   profile variations in a sample of 13 Sgs with spectral types ranging from O9
   to B8.

The letter is structured as follows: the sample of stars and the spectroscopic observations used 
for this study are presented in Sect. \ref{observations}; we characterize the line-broadening
and the line-profile variations of photospheric lines in Sects. \ref{broadening} and \ref{signatures}, 
respectively; a connection between the extra broadening and line profile variations is presented 
in Sect. \ref{connection}. Finally, we discuss the possible relation to stellar pulsations in
Sect. \ref{discussion}.

\section{Observational data\,set}\label{observations}

\begin{table*}[t!]
\begin{center}
{\scriptsize
\caption{Columns 1--5: Targets, spectral classification, observing campaign, time\,span of the 
observations, and number of spectra. Columns 6--11: Line-broadening and LPVs characterization 
in the \ion{Si}{3}\,4567 or \ion{O}{3}\,5592 lines for our stellar sample. LPVs characterized 
with the peak-to-peak amplitudes of the first and third moments of the lines. $\Delta$T in 
days; \vsini, $\Theta_{\rm RT}$, $\langle v \rangle_{\rm pp}$, and $\langle v^3 \rangle^{1/3}_{\rm pp}$ in \kms.
Uncertainties in $\langle v \rangle_{\rm pp}$, and $\langle v^3 \rangle^{1/3}_{\rm pp}$ are 
$\sim$0.10-0.30 and $\sim$0.4-1.2 \kms, respectively
}\label{t1}
\centering
\begin{tabular}{lccccccccccccc}
\noalign{\smallskip}
\hline \hline
\noalign{\smallskip}
        &  &  & & & & \multicolumn{2}{c}{\vsini(FT) } & & \multicolumn{2}{c}{$\chi^2$ fitting} & & \multicolumn{2}{c}{LPVs} \\
\cline{3-5} \cline{7-8} \cline{10-11} \cline{13-14}
\noalign{\smallskip}
Star    & SpT \& LC & Run   & $\Delta$T & \# & &Range & Median & & $\langle$\vsini$\rangle$ & $\langle\Theta_{\rm RT}\rangle$ & & $\langle v \rangle_{\rm pp}$ & $\langle v^3 \rangle^{1/3}_{\rm pp}$  \\
\hline
\noalign{\smallskip}
HD\,207198 & O9\,Ib-II & FI09 & 3.14 & 14 & & 59--64  & 62$\pm$2   & & 50$\pm$2   & 100$\pm$0  & &  3.41 &  26.9 \\
HD\,209975 & O9.5\,Iab & FI08 & 3.04 & 21 & & 56--62  & 59$\pm$2   & & 50$\pm$6   & 95$\pm$5   & & 10.13 &  46.3 \\
           &           & FI09 & 3.18 & 18 & & 52--58  & 57$\pm$2   & & 51$\pm$4   & 100$\pm$5  & & 11.65 &  45.4 \\
HD\,37742  & O9.7\,Ib  & FI09 & 3.19 & 30 & & 90--140 & 117$\pm$10 & & 115$\pm$15 & 117$\pm$13 & & 12.01 &  62.8 \\
HD\,204172 & B0\,Ib    & FI09 & 3.14 & 14 & & 38--68  & 58$\pm$11  & & 58$\pm$6   & 85$\pm$5   & & 11.80 &  44.0 \\
HD\,37128  & B0\,Ia    & FI08 & 3.01 & 60 & & 45--64  & 55$\pm$9   & & 57$\pm$8   & 90$\pm$6   & & 12.30 &  43.0 \\
           &           & FI09 & 3.18 & 28 & & 32--60  & 49$\pm$8   & & 54$\pm$9   & 90$\pm$9   & & 13.02 &  46.5 \\  
HD\,38771  & B0.5\,Ia  & FI08 & 2.99 & 48 & & 42--57  & 50$\pm$6   & & 48$\pm$5   & 90$\pm$6   & &  9.38 &  36.6 \\
           &           & FI09 & 3.17 & 25 & & 30--55  & 48$\pm$8   & & 52$\pm$7   & 90$\pm$5   & & 10.56 &  39.1 \\  
HD\,2905   & BC0.7\,Ia & FI08 & 3.07 & 30 & & 44--58  & 53$\pm$5   & & 51$\pm$6   & 90$\pm$3   & & 10.07 &  39.8 \\  
           &           & FI09 & 3.29 & 28 & & 33--65  & 49$\pm$11  & & 61$\pm$2   & 85$\pm$5   & & 11.49 &  39.6 \\  
HD\,24398  & B1\,Iab:  & FI09 & 3.19 & 34 & & 27--43  & 34$\pm$5   & & 38$\pm$2   & 65$\pm$0   & &  3.15 &  25.7 \\
HD\,190603 & B1.5\,Ia$^+$ & FI08 & 1.96 & 17 & & 30--47  & 46$\pm$5   & & 49$\pm$4   & 55$\pm$3   & &  2.18 &  24.6 \\
HD\,14818  & B2\,Ia    & FI08 & 3.00 & 16 & & 32--47  & 41$\pm$6   & & 41$\pm$5   & 70$\pm$3   & &  7.84 &  37.1 \\
HD\,206165 & B2\,Ib    & FI09 & 3.14 & 16 & & 29--41  & 38$\pm$2   & & 38$\pm$2   & 60$\pm$0   & &  5.63 &  28.2 \\
\noalign{\smallskip}										
HD\,191243 & B5\,Ib    & FI09 & 2.10 & 9  & & 15-25   & 18$\pm$3   & & 18$\pm$2   & 30$\pm$0   & &  1.36 &  13.4 \\
HD\,34085  & B8\,Iab:  & FI09 & 3.08 & 21 & & 14-33   & 23$\pm$4   & & 29$\pm$4   & 35$\pm$5   & &  6.18 &  19.8 \\
\noalign{\smallskip}										
HD\,214680 & O9\,V     & FI08 & 3.12 & 31 & & 13--23  & 18$\pm$3   & & 16$\pm$2   & 40$\pm$5   & &  2.12 &  17.4 \\
HD\,37042  & B0.5\,V   & FI08 & 2.92 & 20 & & 32--34  & 33$\pm$1   & & 33$\pm$0   & 15$\pm$5   & &  0.84 &  10.4 \\
\hline \hline
\end{tabular}
}
\end{center}
\end{table*}

We obtained time series spectra of a selected sample of 11 bright late-O 
and early-B Sgs during two observing runs (2008/11/05-08, 2009/11/09-12)
with the FIES cross-dispersed high resolution echelle spectrograph on 
the 2.5m  Nordic Optical Telescope at  Roque de los Muchachos Observatory on La Palma (Canary 
Islands, Spain). We used FIES in the medium resolution mode (R=46000, 
$\delta\lambda$=0.03 \AA/pix). The sample was complemented with two OB dwarfs
and two late B Sgs. The exposure times were chosen so as to reach at least 
 SNR=200 (measured in the range 4500\,--\,4600 \AA). The list of observed stars, 
along with their spectral classification, number of spectra obtained for each target, 
time\,span of the corresponding time-series, and observing run in which they 
were observed are indicated in the first columns of Table \ref{t1}.
Note that four of these stars were observed in both campaigns.

The spectra were reduced with the FIEStool\footnote{http://www.not.iac.es/instruments/fies/fiestool/FIEStool.html} 
package in advanced mode. A proper set of bias, flat and arc frames obtained 
each night was used to this aim. The FIEStool pipeline provided 
wavelength-calibrated, blaze-corrected, order-merged spectra of high quality. Next, the 
barycentric correction was performed. We used the interstellar \ion{Na}{1}\,D\,doublet 
at 5890 \AA\ to check the accuracy in this correction, and found an agreement 
in the line doublet position for all the time series spectra of each star better 
than 0.5 \kms. 

Once the lines of interest for this study were identified, a local normalization
of selected regions of the spectra was performed. For each 
line, the local continuum used for the normalization was calculated from a linear fit between points inside two 
continuum regions adjacent to the line. The same continuum windows (selected manually 
in the average spectrum) were used for all the spectra of a given star.

\section{Characterization of the line broadening in photospheric lines}
\label{broadening}

Several types of broadening mechanisms are traditionally considered to
produce the final observed line profiles of photospheric metal lines
in massive stars: instrumental, natural, thermal, microturbulent, macroturbulent
and rotational broadening. In the case of high resolution\footnote{
The instrumental broadening contribution can be minimized by using high resolution 
spectra. In our FIES@NOT data\,set (R=46000) the instrumental Gaussian profile 
has a FWHM $\sim$6.5 \kms.} spectra of late-O and 
early B\,Sgs, rotational and macroturbulent constributions dominate.  

We used the Fourier transform technique (\citeauthor{Gra76} \citeyear{Gra76}; see 
also \citeauthor{Sim07}, \citeyear{Sim07} for a recent application to the spectra 
of O and B stars) to disentangle the contributions from rotational and 
{macroturbulent} broadening. 
The results of the analysis are presented in Table \ref{t1}. We performed
the analysis for each of the time series spectra, obtaining  \vsini\ as indicated
by the first zero of the Fourier transform. The range and median of derived
\vsini\ values are indicated in columns 6 and 7 of Table \ref{t1}. Note that the
dispersion in the obtained \vsini\ values is between 10\% and 30\%, depending on the star.
Whether this dispersion is real or an effect of noise is not clear from this
data\,set. As outlined by \citet{Sim07}, the correct identification of the first 
zero is complicated in cases of low SNR and for a large 
 extra broadening contribution. This will be explored in more detail in a future investigation.

Next, we applied a {goodness-of-fit} method \citep[viz.][]{Rya02} to quantify the 
contribution of the extra broadening in each spectrum from the time series. We considered
as free parameters the equivalent width, the radial velocity, \vsini, and the parameter 
defining the extra broadening. We allowed  \vsini\ to vary in the fitting process
in order to compare with results from the FT analysis. Two possibilities for the 
characterization of the extra broadening were considered: an isotropic Gaussian 
profile, 
and a Gaussian radial--tangential one \citep[$\Theta_{\rm RT}$, 
see][]{Gra76}. In the latter, we assumed that the radial and tangential 
components provide equal contributions 
to the final profile. The analysis using an isotropic Gaussian profile for the {macroturbulent}
broadening resulted in
(a) a large dispersion in the \vsini\ values obtained from the various 
spectra in each time series; (b) systematically lower \vsini\ values than those 
determined through the FT method (a factor of 0.5 in many cases, sometimes even ``zero'' 
rotation) for all stars with dominant extra broadening. On the other 
hand, when assuming a Gaussian radial--tangential profile, we found that (a) the 
dispersion in the \vsini\ and $\Theta_{\rm RT}$ values was rather low; (b) differences 
between the \vsini\ values obtained from the {goodness-of-fit} method and using the FT 
technique are $\le$\,$\pm$20\%. We thus conclude that a Gaussian radial--tangential profile 
is better suited to characterizing the extra line broadening than an isotropic Gaussian 
profile.\footnote{We also found that the former generally produces a better final fit 
to the observed profiles, though the differences are subtle.} The corresponding mean 
and standard deviation for \vsini\ and $\Theta_{\rm RT}$ as obtained from the $\chi^2$
fit are indicated in columns 8 
and 9 in Table \ref{t1}. Note the good agreement in the derived values for the four 
stars observed in both  2008 and 2009.

Similarly to previous studies \citep[e.g.][]{Duf06, Lef07, Mar08, Fra10}, we found 
that the size of the extra broadening is generally larger than the rotational 
contribution in all the Sgs. This is not 
the case for the B0.5\,V star HD\,37042, where the total broadening is mainly 
produced by the effect of the stellar rotation
(note, however, that a certain extra broadening is also needed in this case). 
The other dwarf star, HD\,214680, is a special case, since it has very low \vsini. 
For such a low \vsini\ values, microturbulence produces a significant contribution to the 
total broadening, which then is included in the measured extra broadening. Note that this
argument may also explain the extra broadening needed for HD\,37042 (and the other
two late-B Sgs with low \vsini).

\section{Characterization of line profile variations in photospheric lines}
\label{signatures}

   \begin{figure*}[t!]
   \centering
   \includegraphics[width=5.5cm, angle=90]{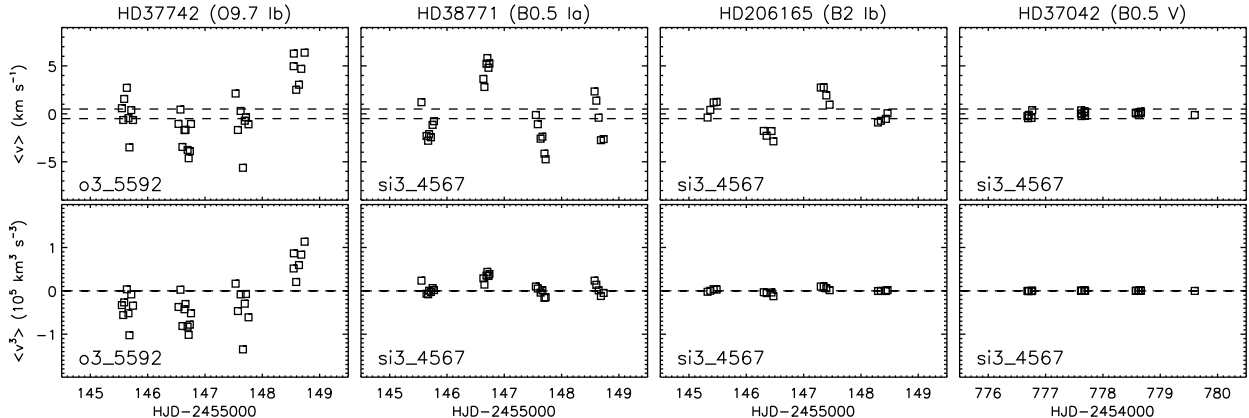}
      \caption{Representative examples for the variation of the first moment (radial velocity placed 
      at average zero) and third moment (skewness of the line) for some of the stars in the studied
      sample derived from the \ion{Si}{3}\,4567 or \ion{O}{3}\,5592 line profiles. Horizontal 
      lines in upper plots show the accuracy in radial velocity associated with the instrumental 
      setting used for FIES@NOT.}
         \label{f1}
   \end{figure*}

In agreement with earlier studies of spectroscopic variabi\-lity in
O and B Sgs (e.g. \citealt{Ebb82}; \citealt{How93}; \citealt{Ful96}; 
\citealt{Pri96}, \citeyear{Pri04}, \citeyear{Pri06}; \citealt{Mor04}; 
\citealt{Kau06}; \citealt{Mar05}, \citeyear{Mar08}), we
found clear signatures of line profile va\-ria\-tions (LPVs) for all the
Sgs considered in our sample.
To  investigate these LPVs quantitatively we 
computed the first, $\langle v \rangle$, and 
third, $\langle v^3 \rangle$, normalized velocity 
moments \citep[for definitions, see, e.g.,][]{Aer10a}
from the \ion{Si}{3}\,4567 or \ion{O}{3}\,5592 
lines with FAMIAS\footnote{A software package developed 
in the framework of the FP6 European Coordination Action HELAS 
(http://www.helas-eu.org/).} \citep{Zim08}. These moments are related to the centroid 
velocity and the skewness of the line profile, respectively, and are 
well suited for deciding whether an observed line profile is subject to 
time-dependent line asymmetry (note that their values are zero for purely symmetric 
profiles).

Four representative examples of the temporal behavior of the first and third
velocity moments are presented in Fig.\,\ref{f1}. The associated uncertainties
(not included in the plot) are $\sim$\,0.1\,-\,0.4 \kms, and
$\sim$\,1000\,--\,4000 km$^3$\,s$^{-3}$, respectively.  In the case of the dwarf
star HD\,37042, the $\langle v \rangle$ values are fairly constant and close to
zero. The maximum dispersion in velocity for this star is $\sim$1 \kms, of the
order of the accuracy associated with the instrumental setting used for the
FIES@NOT observations (indicated as dashed ho\-ri\-zon\-tal lines). All the
other stars in Figure \ref{f1} show $\langle v \rangle$ variations above this significance level,
with maximum peak-to-peak amplitudes reaching $\sim$10--12 \kms\ in some of the cases. 
The minimum variations are found for
HD\,214680 (the other luminosity class V object con\-si\-de\-red in this study),
HD\,191243 (B5\,Ib) and HD\,190603 (B1.5\,Ia$^+$), with maximum amplitudes
slightly larger than the significance level.  Curiously, the star HD\,190603,
expected to be an extreme object from its spectral classification, is one of the
cases with smaller variations.  We indicate in Table \ref{t1} (columns 10 and
11) the peak-to-peak amplitude of the first and third moment variations measured
for each of the considered targets.

\section{The {``macroturbulence''}-LPV connection}\label{connection}

\begin{figure}[t!]
   \centering
   \includegraphics[width=8cm, angle=90]{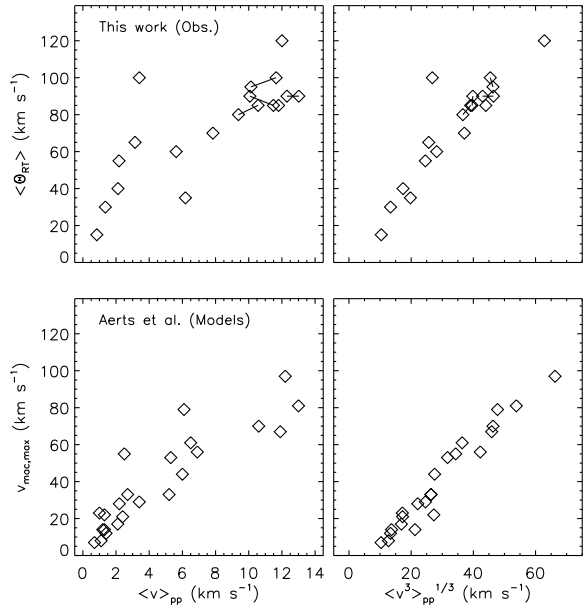}
      \caption{(Top) Empirical relations between the average size of the macroturbulent broadening
	  ($\langle\Theta_{\rm RT}\rangle$) and the peak-to-peak amplitude of the first and third moments of the
	  line-profile.(Bottom) Similar relations are found from the simulations by \cite{Aer09}.}
         \label{f2}
   \end{figure}

We investigate the possible connection between {macroturbulent} broadening
and LPVs in our sample of stars. Such a connection should appear in case macroturbulent 
broadening is produced by any time-dependent physical 
phenomena \citep[e.g. non-radial oscillations, see][]{Aer09}. In Fig.\,\ref{f2} (upper pannels), 
we plot the average size of the {macroturbulent} broadening, $\langle \Theta_{\rm RT} \rangle$, versus 
the peak-to-peak amplitude of $\langle v \rangle$ and $\langle v^3 \rangle$ variations
for each of the  stars studied. Results from the four stars observed in both campaigns
are conected with solid lines. A clear positive correlation is present in both cases.  
To our knowledge, this is the {\em first clear observational evidence for a 
connection between extra broadening and LPVs in B and late O Sgs.}   
Particularly remarkable is the $\Theta_{\rm RT}$\,-\,$\langle v^3 \rangle$ correlation: 
the larger the extra broadening, the more asymmetric line profiles can be found in the 
time series. Note that this does not mean that lines with a significant {macroturbulence} 
contribution are always asymmetric since $\langle v^3 \rangle$ oscillates between positive 
and negative values over time.

We also present in Fig.\,\ref{f2} (bottom pannels) similar plots with data
  from Table 1 in \cite{Aer09}, based on simulations of line profiles broadened
  by rotation and by hundreds of low amplitude non-radial gravity mode
  pulsations. These simulations consider various combinations of the inclination
  angle ($i$), the projected rotational velocity (\vsini), and the intrinsic
  amplitude of the modes denoted by $a$ \citep[see][for a definition]{Aer09}.  We
  present the maximum values of v$_{\rm mac}$ obtained for each set of simulated
  profiles. Similar trends for $v_{\rm mac}$ as in Fig.\,\ref{f2} are found
  using average or minimum values for the macroturbulence parameter, but with a
  different (smaller) scale in the y-axis.

The simulations lead to clear trends which are compatible with spectroscopic
observations.

\section{Is macroturbulent broadening in OB-Sgs caused by pulsations?}
\label{discussion}
Non-radial oscillations have been often suggested as the origin of LPVs and
photosperic lines in OB Sgs, as well as the driver of large scale wind
structures (see references in Sect.\,\ref{signatures}); however, a firm
confirmation (by means of a rigorous seismic analysis) has not been achieved
yet. From a theoretical point of view, g-mode oscillations were not initially
expected in B\,Sgs because the radiative damping in the core was suspected as
being too strong. \citet{Sai06} detected simultaneous p- and g-modes in
HD\,163899 (B2\,Ib/II) using data from the MOST satellite. These authors also
computed new models showing that g-modes can be excited in massive post-main
sequence stars, as the g-modes are reflected at the convective zone associated
with the H-burning shell. \citet{Lef07} presented observational evidence of
g-mode instabilities in a sample of photometrically variable B\,Sgs from the
location of the stars in the (log\,$g$, log\,T$_{\rm eff}$)-diagram.

These results, along with our observational confirmation of a tight
connection between {macroturbulence} and 
parameters describing observed LPVs render stellar oscillations the
most probable physical origin of macroturbulent broadening in B\,Sgs; however,
it is too premature to consider them as the only physical phenomenon to explain
the unknown broadening. 

Recent studies of high resolution spectra of OB stars indicate that the
appearance of a significant macroturbulent broadening contribution is not only
limited to B Sgs but occurs also in O stars of all luminosity classes \citep{Sim10, Mar10}
as well as
in main-sequence B-type pulsators \citep[e.g.][]{Mor06}. Moreover, with the
recent discovery of a strange-mode oscillation in the B6Ia supergiant HD\,46005
\citep{Aer10b} and of stochastic oscillations in the O8V star HD\,46149
\citep{Degr10} and the $\beta$\,Cephei star V1449\,Aql \citep{Bel09} from high-precision 
space photometry, observational evidence of atmosphere phenomena due to oscillations, which
{\it must\/} cause some level of pulsational broadening, is growing.

Future investigations of the correlation between LPVs and macroturbulence, and
its connection to stellar oscillations in the whole realm of massive stars, will
be key to establishing the role played by the different physical processes
contributing to the spectral line formation in these types of stars. Different 
models including various combinations of pulsational, rotational and wind effects 
should be considered, keeping in mind that an appropriate physical mechanism 
must be able to explain the observed correlation presented in this study.

\acknowledgments

{Financial support from the Spanish Ministerio de Ciencia e Innovaci\'on under
  the project AYA2008-06166-C03-01 is acknowledged.  This work has also been
  partially funded by the Spanish MICINN under the Consolider-Ingenio 2010
  Program grant CSD2006-00070: First Science with the GTC
  (http://www.iac.es/consolider-ingenio-gtc). The research leading to these
  results has also received funding from the European Research Council under the
  European Community's Seventh Framework Programme (FP7/2007--2013)/ERC grant
  agreement n$^\circ$227224 (PROSPERITY).}

\end{document}